# Local Symmetry Breaking Drives Picosecond Spin Domain Formation in Polycrystalline Semiconducting Films

Arjun Ashoka[1], Satyawan Nagane[2], Nives Strkalj[3], Bart Roose[2], Jooyoung Sung[4], Judith L. MacManus-Driscoll[3], Samuel D. Stranks[2], Sascha Feldmann[5*], Akshay Rao[1*]

[1]Cavendish Laboratory, University of Cambridge, J.J. Thomson Avenue, Cambridge, CB3 0HE, United Kingdom

[2]Department of Chemical Engineering and Biotechnology, University of Cambridge, Philippa Fawcett Drive, Cambridge CB3 0AS, United Kingdom

[3]Department of Materials Science and Metallurgy, University of Cambridge, 27 Charles Babbage Road, Cambridge CB3 0FS, United Kingdom

[4]Department of Emerging Materials Science, DGIST, Daegu, 42988 Republic of Korea

[5]Rowland Institute, Harvard University, Cambridge, MA 02142, USA

Correspondence: sfeldmann@rowland.harvard.edu, ar525@cam.ac.uk

**Abstract:**

Photoinduced spin-charge interconversion in semiconductors with spin-orbit coupling could provide a route to optically addressable spintronics without the use of external magnetic fields. A central question is whether the resulting spin-associated charge currents are robust to structural disorder, which is inherent to polycrystalline semiconductors that are desirable for device applications. Using femtosecond circular polarization-resolved pump-probe microscopy on polycrystalline halide perovskite thin films, we observe the photoinduced ultrafast formation of spin-polarized positive and, unexpectedly, negative spin domains on the micron scale formed through lateral currents. Further, the polarization of these domains and lateral transport direction is switched upon switching the polarization of the pump helicity. Micron scale variations in the intensity of optical second-harmonic generation and vertical piezoresponse suggest that the spin domain formation is driven by the presence of strong local inversion symmetry breaking via inter-grain structural disorder. We propose that this leads to spatially varying Rashba-like spin textures that drive spin-momentum locked currents, leading to local spin accumulation. Our results establish ultrafast spin domain formation in polycrystalline semiconductors as a new optically addressable platform for nanoscale spin-device physics.

**One-Sentence Summary:** Polycrystalline disorder leads to local spin order

Central to the use of spin in solid state spintronic devices is the ability to lift the energetic degeneracy of spin polarized electrons - Kramers degeneracy. Historically this has been achieved by manipulating the local magnetic environment of electrons through an external magnetic field, which breaks time-reversal symmetry and lifts Kramers degeneracy (*1*). This degeneracy can also be lifted without the need for an external magnetic field in systems with spin-orbit coupling (SOC) that break inversion symmetry – leading to spin and momentum eigenvalues that are locked to each other. This is known as either the Bychkov–Rashba effect on the surface or Dresselhaus effect when present in the bulk (collectively referred to as the Rashba effect henceforth) (*2*, *3*).

With inversion symmetry broken in the out-of-plane direction, a radially symmetric spin-orbit perturbation splits the initially degenerate spin bands in k-space (Fig. 1A). The two bands that emerge are no longer eigenstates of in-plane spin and they inherit particular k-dependent directions referred to as spin-textures (red and blue arrows forming circles in Fig. 1A). Along any cut in the direction perpendicular to the symmetry breaking, the inner and outer bands on opposite sides of the zone center exhibit the same spin-state, labelled by the spin-quantum number, $m_j$ (green or orange bands, Fig. 1A) (*4*, *5*). Upon photoexcitation with circularly polarized light from a similarly split valence band, selection rules dictate that $\Delta m_j = \pm 1$ (depending on the light helicity), populating one semicircle of the Fermi contour of the inner spin texture, and the zone-opposed semicircle of the Fermi contour of the outer spin texture (*5*). This leads to a net photoinduced electronic momentum imbalance as the Fermi wavevectors for each spin texture is different (*6*). From standard Boltzmann transport treatment, this manifests as an in-plane charge current (*6*–*9*). Depending on the specific dimensionality and chirality of the spin induced charge current, this effect is known as the inverse spin Hall, inverse Rashba Edelstein, circular photogalvanic or anomalous circular photogalvanic effect (*7*, *10*–*13*). In all these effects, the link between the particular selection rule (helicity of light) and associated charge current direction is dictated by the spin textures – if the spin textures were exchanged, the charge current would reverse direction.

**Results**

Here we study spin currents in polycrystalline films halide perovskites, which show high quality optoelectronic properties. We focus on 150 nm thin films of the high-performing alloyed composition $FA_{0.79}MA_{0.16}Cs_{0.05}Pb(I_{0.83}Br_{0.17})_3$ (where FA is formamidinium and MA is methylammonium), which have a high degree of polycrystallinity (typical reported grain sizes ranging from 50 to 1000 nm) and a bandgap at 764 nm (see Fig. S1 for absorption and luminescence spectra) (*19*).

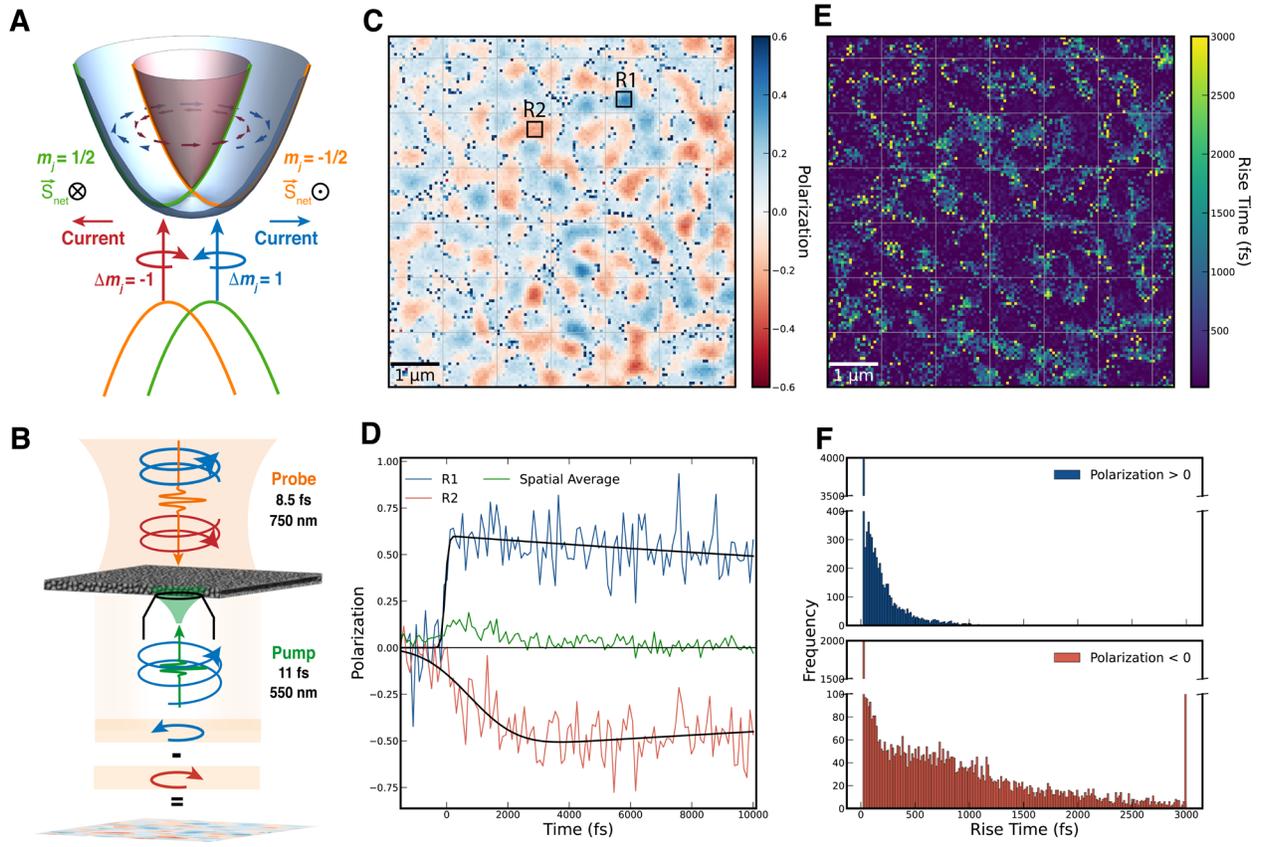

**Fig 1: A.** Schematic of the effect of spin-selective photoexcitation on induced lateral charge currents in Rashba-like bands. **B.** Schematic of the experiment where a circularly polarized broadband 11-fs pump pulse excites a large region of the sample and a circularly polarized (co (blue) or counter (red)) probe pulse subsequently images the sample after a variable time delay. The resulting images are recorded and the polarization image (co minus counter polarized probe) are obtained. **C.** An image of the maximum intensity of the photoinduced circular polarization shows regions of both negative and positive polarizations, with two example regions (R1, R2) marked. **D.** The kinetics of the regions shown in C with positive polarization (R1, blue) display an ultrafast rise (sub 100-fs), while those with negative polarization (R2, red) show a delayed rise of a few ps with fits shown as black lines. The spatial/ensemble average exhibits a fast decay consistent with previous literature (green). **E.** Upon fitting the kinetics at every pixel in C, we find that the regions with a delayed rise are strongly correlated with the regions of negative polarization, as is clearly demonstrated by the histograms in **F**.

To resolve the real space propagation of spin-currents and study the effects of the spatial inhomogeneity inherent to these films on the local spin depolarization, we perform ultrafast circular polarization-resolved pump-probe imaging in a widefield microscope geometry at room temperature. The principle of circular polarization-resolved pump-probe measurements is as follows: the studied system is first photoexcited with a pump pulse of a given helicity ($\Delta m_j = +1$ or -1, green line, blue/red arrows Fig. 1B), resulting in a net polarization of electronic spin. After a fixed time delay, a probe pulse (orange Fig. 1B), of the same (co, $\Delta m_j = +1$ or -1) or opposite (counter, $\Delta m_j = -1$ or +1) helicity is incident on the material, and the difference between the transmission of the probe pulses ($\Delta T/T_{co} - \Delta T/T_{counter}$) due to the pump pulse is reported as the photoinduced circular polarization. In halide perovskite films, it is reported that this net ensemble positive photoinduced spin polarization decays to zero in a few picoseconds due to spin scattering events that randomize the spins of the electrons (*20*, *21*).

We first photoexcite a widefield region of the film ~ 200 μm², with a 11-fs pump-pulse (500-650 nm,) and probe a 100 μm x 100 μm region of this sample with an 8.5 fs broadband probe-pulse (700-900 nm) (see Fig. S2 -S6 for pulse compression, temporal and polarization characterization, Fig. S7 - S9 for sweep-to-sweep spatially averaged kinetics). We study a homogeneously excited region of 7 μm x 7 μm, choosing to image at a wavelength of 750 ± 10-nm, near the bandgap, where there is a large ground-state bleach signal and where previous studies on spin-lifetimes in perovskites have focused (*20*). We label right (left) circularly polarized light as $\sigma$ + (-). To study the spatial photoinduced circular polarization signal, we subtract widefield ΔT/T images of the probe co and counter polarized to the pump helicity at each time delay and report this as the local spin polarization image (see Fig. S10 for widefield $\Delta T/T_{co}$ and $\Delta T/T_{counter}$ images). As we make no comment on the polarization magnitude in this report and focus solely on the kinetics and signal sign, we normalize the signals to the average image signal to get an estimate for the magnitude (see SI text 1 for discussion on avoiding linear dichroism and birefringence artefacts). The reported polarization is thus calculated as,

$$\text{Polarization } (x,y,t) = \frac{\Delta T/T_{co}(x,y,t) - \Delta T/T(x,y,t)_{counter}}{\langle \Delta T/T(x,y,t)_{co} + \Delta T/T(x,y,t)_{counter} \rangle_{x,y,t}}. \tag{1}$$

In order to extract local information on the polarization dynamics (see Movie S1 for video), we fit every pixel to a Gaussian-convolved exponential decay (see SI text 2 for equation).

Fig.1C shows that, upon fitting the kinetics of the measured widefield excited polarization images, we retrieve large spatial variations in the sign of the polarization signal, with regions of positive and negative polarization. This is surprising as a net counter-polarized signal is not expected to occur based on the previously reported spin-dynamics of these systems (*20*). If we average over all spatial points, we do indeed retrieve the previously reported ensemble average behavior, as shown in Fig.1D (green line) (*20*). Considering all points individually however, we observe a binary behavior, with domains of positive polarization rising immediately upon photoexcitation and domains of negative polarizations showing a delayed risetime over a few picoseconds, as exemplified for domains R1 and R2 in Fig.1C with corresponding kinetics and fits shown in Fig.1D. We note that both types of domains show longer than spatially averaged spin lifetimes (green line in Fig 1D, Fig. S7-S9), more consistent with reports of single crystal spin lifetimes (*22*).

Fig. 1E shows the rise time for each point within the field of view. A clear correlation between the sign of the polarization (show in Fig. 1C) and its associated rise time (Fig. 1E) can be seen, with domains of negative polarization rising later than domains of positive polarization. This is more clearly seen in the histograms in Fig. 1F. In the top panel of Fig. 1F we note that there is a spread in rise times, which comes from locally varying carrier cooling rates as we photoexcite high in the band (see Fig. S11 for probe energy dependence) (*23*). However, when compared to the bottom panel, the timescale separation for the positive and negative polarizations is well above this broadening. We repeat this on films prepared by a different synthetic route of different thickness (100-nm) and find the same effect (see Fig. S12). On repeating the same experiments with linear polarized pump and probe pulses we observe no correlation between the local rise times and sign of polarization anisotropy (see SI text 1, Fig. S13 and S14).

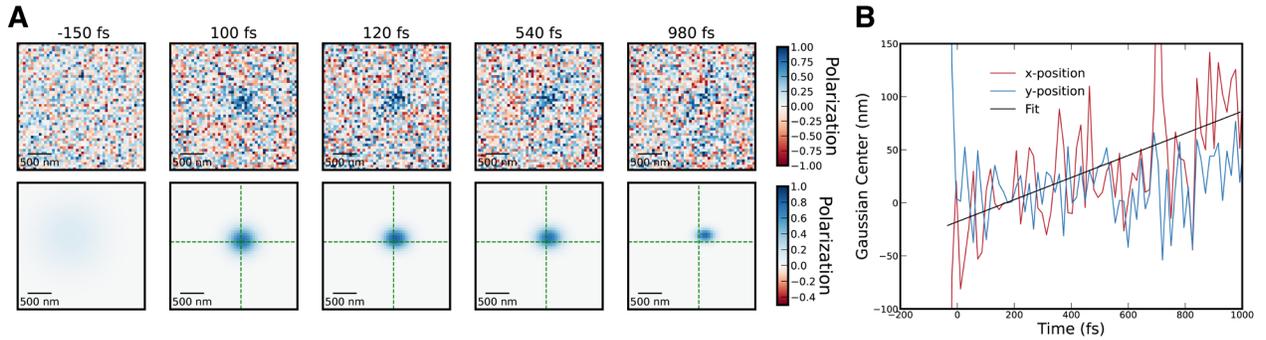

**Fig 2: A.** Images of the spatial spin-polarization induced by a focused polarized pump pulse. The images show a clear lateral movement of the polarization away from its initial position at 100 fs (green dashed lines) over 1 ps, with the bottom images showing a Gaussian fit to the signal. **B.** Results of a Gaussian fit to the polarization show a lateral movement in the center along the x-coordinate of 100 nm in 1 ps.

To study the origin of this rise-time-correlated negative polarization signal, we investigate the spatial transport dynamics of the photoinduced polarization (Fig. 2). Here we perturb the sample with the same circularly polarized pump pulse, but now focused through the microscope objective to a near diffraction-limited 200-nm spot. This photoinduces a local Gaussian spin-polarized electronic population, the transport of which is then imaged with the co and counter polarized widefield probe pulse as described earlier. To study the motion of the spin-polarized electrons, we directly fit a Gaussian function to the polarization image and study the difference between co and counter polarized probe images. As seen in Fig. 2A, we measure a Gaussian distributed net polarization at early times (100 fs). Over the first picosecond however, we find that, rather than expanding isotropically as might be expected, the photoinduced polarization moves away from the initial spot (as seen in the images at 120 fs, 540 fs and 980 fs) (*24*). We repeat this on several spots and different excitation densities and uniformly observe this behavior (Fig. S15 and S16). Fitting the change in the Gaussian center of the net polarization signal as a function of time, we resolve motion of 100 nm over 1 ps, yielding a velocity of about $10^5$ m/s (Fig. 2 B). This velocity is likely related to the group velocity of the carriers near the band edge convolved with the Fermi velocity which sets the drift velocity of spin-momentum locked currents in Rashba-like bands (*24*).

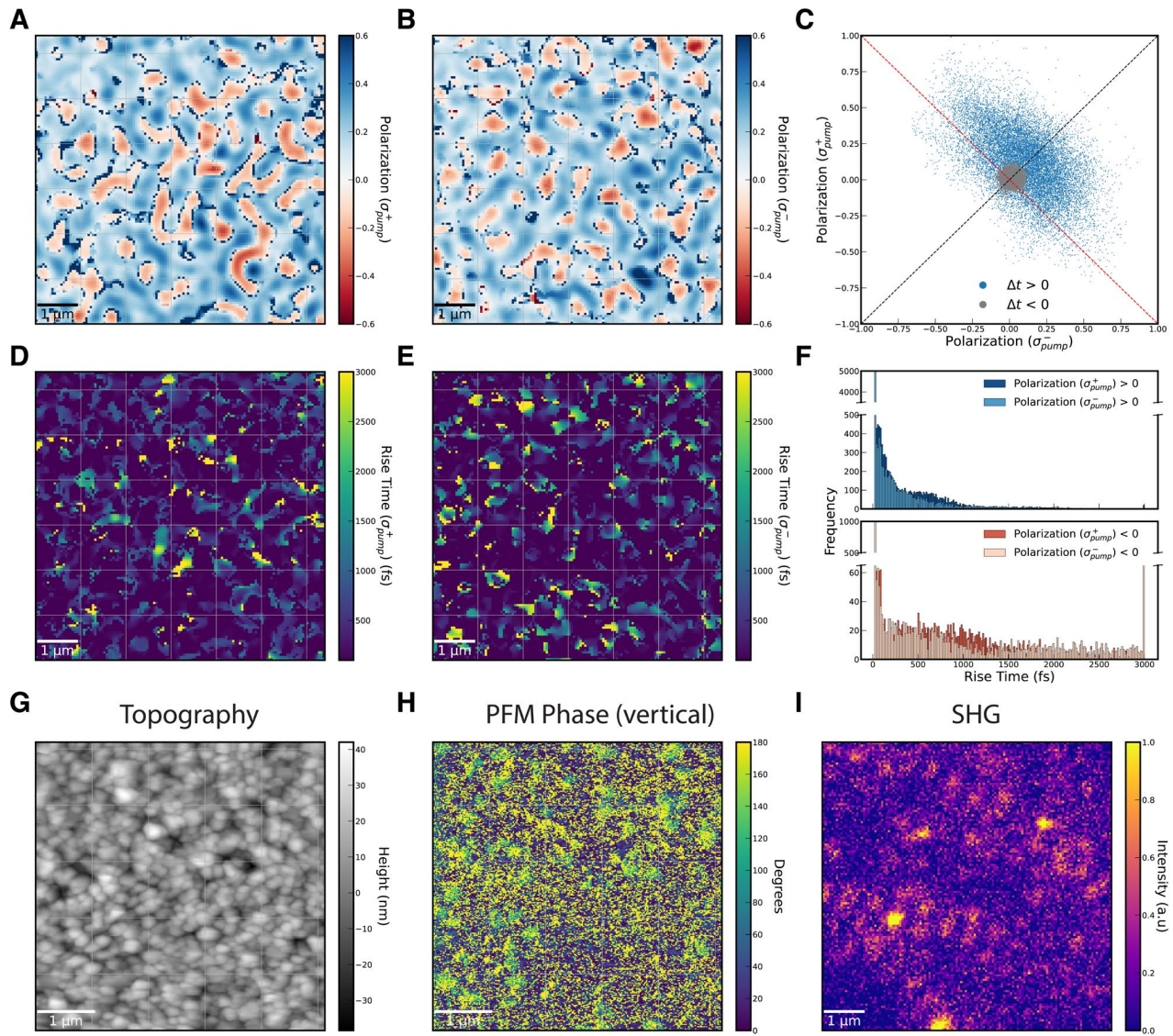

**Fig 3: A.** Photoinduced circular polarization upon right-handed (σ +) polarized pump. **B.** Photoinduced polarization upon left-handed (σ -) polarized pump shows domains of inverted sign when compared to A. **C.** Photoinduced polarization signal at each pixel (blue dots) before and after the arrival of the pump pulse, showing a clear spatial anticorrelation in the sign of the signal with pump helicity after the pump pulse arrives. **D.** Just as in Fig 1, we see that the regions of long, picosecond rise times are correlated with the regions of negative signal in A. **E.** On reversing the helicity of the pump to σ -, these regions are still correlated with the negative regions in the polarization. **F.** Histograms of the rise time for regions of positive (top) and negative (bottom) polarized regions, which always show longer times for negative electronic polarization, regardless of the pump helicity. **G.** Topography measured using atomic-force microscopy shows micron scale polycrystalline grains in the studied thin films. **H.** The vertical piezoresponse phase shows regions with a 180-degree shift in phase over length scales similar to the grains observed in the topography image, suggesting the presence of polar grains in the film. **I.** This local inversion symmetry breaking is also manifested in the large spatial variations in SHG signal on similar length scales.

This fast, lateral motion suggests that photoinduced spin polarization leads to a lateral current on sub-1 μm length scales and could explain the spatially varying rise of the anomalous negative polarization. There are two possible scenarios which could explain our observation. The first is the motion of spin-induced lateral currents that arise from Rashba-like spin textures. We note that such spin pumped lateral charge currents have been previously reported in topological insulators and bulk single crystal semiconductors using contact based current measurements (*7*, *8*, *14*). The second is apparent spin transport associated with spin-polarized charges moving in a locally varying energy landscape (*25*). As shown in Fig. S17 – S21, we observe a lack of correlation of the measured domains with the spatial photoluminescence as might have been expected for energy driven transport. To further distinguish between these two scenarios, we perform widefield experiments for both pump helicities. If the spin transport is caused by charge motion in a locally varying energy landscape, the direction of spin transport will be invariant to the pump helicity. In contrast, in the Rashba-like scenario of spin-associated currents (Fig. 1A), the direction of spin transport must reverse when the pump helicity is inverted.

As shown in Fig. 3A and 3B, we observe that the local domains of negative and positive polarization reverse sign on changing the helicity of the pump pulse. As expected however, the sign of the spatially averaged polarization decay depends only on the relative polarization between the pump and probe, i.e., it is initially positive and decays to zero as previously reported (*20*) (Fig. S22). The spatial anti-correlation in the polarization on changing the helicity of the pump is made clear in Fig. 3C, where, after the pump pulse arrives ($\Delta t > 0$), the anticorrelation in the signal at each pixel is evident. Crucially, on fitting the kinetics for each pixel, we observe that the domains of slow rise times, that were correlated with the negative polarization, switch with the domains of fast rise time under changing the helicity of the pump pulse, as seen in Fig. 3D and 3E. This shows that the direction of underlying lateral transport is reversed. This timescale separation is also clearly seen in the histograms shown in Fig. 3F, which strongly resemble those in Fig. 1F, regardless of the helicity of the pump pulse.

To gain further insight into the microscopic origins of the local spin-locked currents, we investigate these thin films for signatures of local inversion symmetry breaking, as would be needed to produce local Rashba-like spin textures (*14*). We study the micron-scale polycrystallinity (Fig. 3G) using piezoresponse force microscopy (PFM) and uncover signatures of polarized domains (Fig. 3H, Fig. S23 for corresponding PFM amplitude) (*26*). Fig. 3G and 3H are taken on the same region of the film to correlate the piezoresponse to topography. We are also able to recover switchable ferroelectric loops on some areas of the film (Fig. S24). Local inversion symmetry breaking is further evidenced by micron-scale variations in the SHG intensity measured on the same films as seen in Fig. 3I. We note that while the presence of SHG is evidence for local symmetry breaking, the large variations in intensity could possibly arise from polar domains or the surfaces of locally canted grains.

**Discussion**

Our main observations are fourfold. First, there is a photoinduced ultrafast formation of spin-polarized positive and, unexpectedly, negative spin domains on the micron scale in real space formed through lateral currents. Second, the anomalous domains of negative polarization grow in over a few picoseconds, consistently slower than those with positive polarization which are limited by the instrument response and carrier cooling. Third, the polarization of these domains and lateral transport direction is switched upon switching the polarization of the pump helicity. Forth, there is strong evidence of local inversion asymmetric phases in these polycrystalline films.

In order to explain these observations, we propose a model of spin-momentum locked currents in a micron-scale disordered system. Consider two adjacent regions in space where inversion symmetry is broken in opposite directions (Fig. 4A). The local spin textures in the two regions also have opposite directions of the spin-locked lateral currents for a given pump helicity. Immediately upon photoexcitation of these two regions with right-handed circularly polarized light ($\sigma$ +), these regions in real space are co-polarized (blue). After a few hundred femtoseconds, however, spin-flip mechanisms (Elliott–Yafet or D'yakonov-Perel) cause a subpopulation of the spin-locked currents to reverse in direction (become red). Propagating the spins with their respective current directions leads to the scattering-limited growth of regions of counter-polarized spins (red) and consequently the formation of spatially segregated regions of co- and counter-polarized (blue and red, respectively) spins. In this scheme, the co- and counter- polarized regions reverse sign upon photoexcitation with left-handed circularly polarized light ($\sigma$ -), as do the negatively polarized regions which show the delayed rise due to spin-scattering-limited growth, entirely consistent with our observations.

While this simple model for two regions shown in Fig. 4A works well at capturing the phenomenology of our observations, it is not clear whether this spin-accumulation effect is robust against the polycrystalline nature of our films seen in Fig. 3G. We therefore built a stochastic Monte Carlo based toy model of the spin transport in a polycrystalline, structurally disordered film (see SI text 3 for details). As input parameters to the model we use the measured lateral velocities, spin depolarization lifetimes, symmetry breaking length scales consistent with Fig. 3 and picosecond simulation time scales. As seen in Fig. 4B, each region represents a domain with a random direction and magnitude of inversion symmetry breaking, which results in randomized spin-locked currents. We stress that while this disordered landscape can arise from a multitude of microscopic mechanisms – polar/ferroelectric, orientational/canting, interfacial or strain inhomogeneity – its overall effect is to break inversion symmetry locally. In this disordered landscape, we initialize a uniform distribution of $10^4$ local spin-polarized currents. We then propagate these spin currents in space according to the local spin-momentum locking direction along with the possibility of a Monte-Carlo style random spin flip, which reverses the current direction for a given region (for video see Movie S2, for simulated spin depolarization see Fig. S25). We find excellent agreement between the features observed in our experiment and this model, which are brought out by convolving the modelled data using a Gaussian kernel density estimate (Fig. 4C). This suggests that this toy model captures the observed phenomenology.

While the precise microscopic origins of the local inversion symmetry breaking landscape are not understood to date for the case of halide perovskite materials, we note that recent reports of a local symmetry-lowering octahedral tilt responsible for stabilizing the same perovskite film composition may also be responsible for local inversion symmetry breaking phases, despite reports of an inversion symmetric space group (*27*). Further, strain in halide perovskites has been theoretically predicted to modulate the Rashba parameter and local micron scale strain maps in the same perovskite film composition have reported high uniaxial strain along the grains (*28, 29*). Even if the regions in the film were not bulk-inversion asymmetric, based on previous reports the crystalline grain surfaces are certainly thought to host Rashba bands whose orientation is randomized in a polycrystalline film (*16*). Although this would only give rise to net surface currents, not bulk currents, our model is agnostic to the different spin textures that could be associated to either surface or bulk inversion asymmetry (*5*). Further theoretical work is called for to elucidate the origins of the local Rashba-like spin textures in these materials.

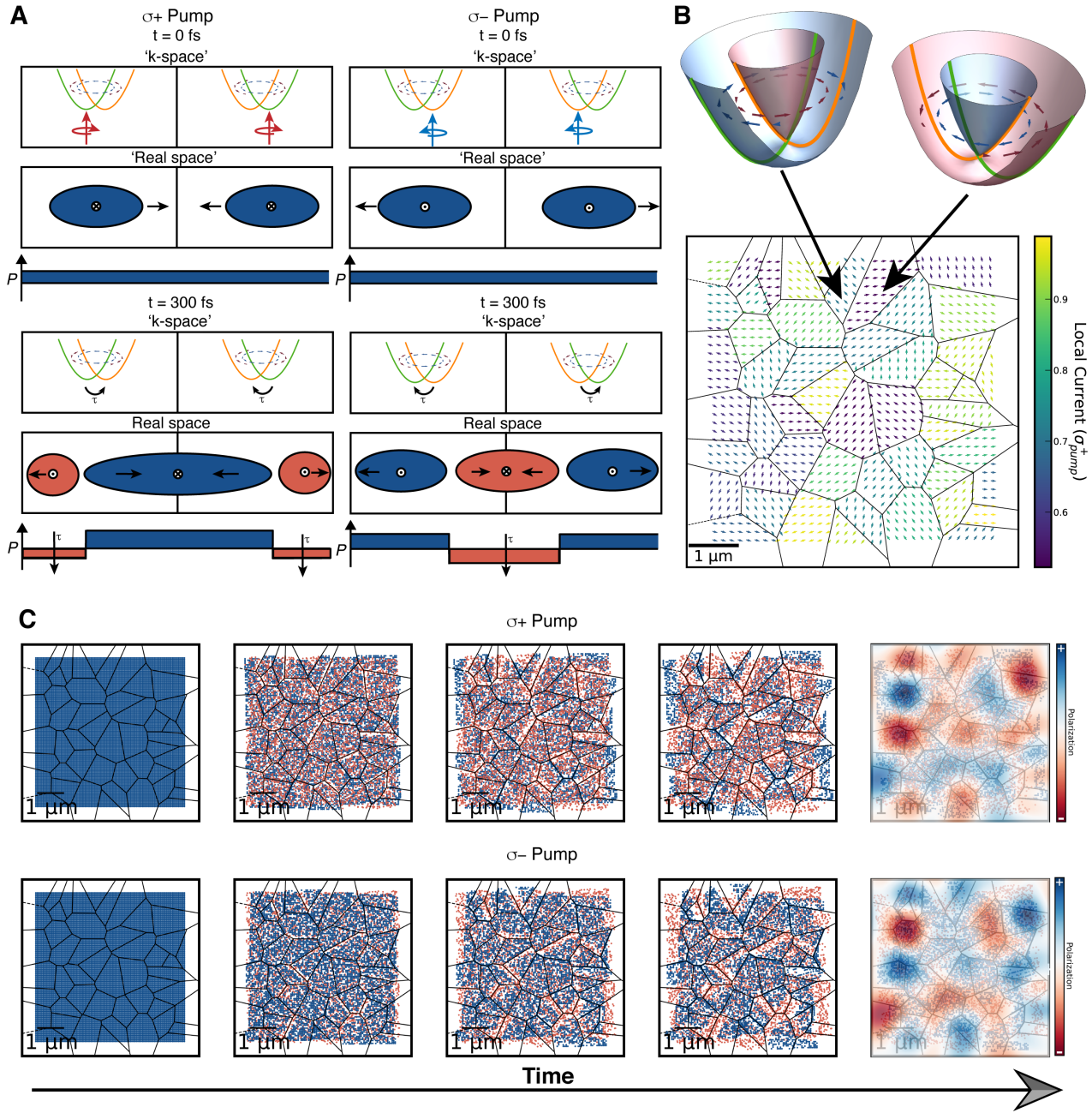

**Fig 4: A.** Two adjacent regions where the directions of inversion symmetry breaking are opposite, the spin textures are switched between the regions, leading to inverted spin-momentum locking, i.e., for a given photoinduced spin, the momentum imbalance direction is inverted. When combined with spin scattering events, this leads to spatial segregation of spins on femtosecond timescales and consequently the scattering-limited growth of negatively polarized domains, which invert sign under inversion of the initial spin population. **B.** Random crystalline domain structures with disorder in the direction and magnitude of inversion symmetry breaking, leading to spatially varying spin-textures and spin-momentum locking. **C.** Propagating initially polarized spin populations using a toy Monte Carlo model to capture stochastic spin flips reveals spatially segregated domains, which when binned used a Gaussian kernel density estimator closely resemble the experimentally uncovered spin polarized domains that invert with inverting the pump helicity.

Finally, we note that regardless of the origin of micron-scale regions with opposite inversion-symmetry breaking (Fig. 4B), our model shows that spin domain formation will always be observed in a system where the local spin current directions are random. For example, if any two neighboring regions in Fig. 4B have a component of their spin-current along the same direction, this necessarily leads to spin-accumulation domains upon spin-flip scattering events. Crucially, we note that these local spin accumulation effects spatially average to zero, which explains why they have remained hidden to bulk experimental probes (optical or contact-based).

Such symmetry-breaking-induced spin accumulation effects on the micron-scale can be utilized to engineer local spin domains by patterning the symmetry-breaking directions through micron scale electric field or elastic strain (for example, see Fig. S24 for ferroelectric hysteresis). This has broad applications such as local spin injection for chiral emission of spinLEDs, and polycrystalline semiconductor analogues to magnetic domain wall logic (*1*, *21*, *30*).

**Acknowledgements:** We thank Alexander J. Sneyd for help with experimental setup development, Ashish Sharma, Emmanuel Gottlob and Richard Friend for helpful discussions.

**Funding**: We acknowledge support from the Engineering and Physical Sciences Research Council (UK) via grants EP/M006360/1 and EP/S030638/1. We acknowledge financial support from the Winton Programme for the Physics of Sustainability. This project has received funding from the European Research Council (ERC) under the European Union's Horizon 2020 research and innovation programme (grant agreement no. 758826). A.A. acknowledges funding from the Gates Cambridge Trust and as well as support from the Winton Programme for the Physics of Sustainability. S.D.N acknowledges the EPSRC (EP/012932/1) for funding. J. D. and N. S. acknowledge funding from the Royal Academy of Engineering Chair in Emerging Technologies grant - CIET1819_24, the Swiss National Science Foundation grant P2EZP2-199913, and the ERC grant EU-H2020-ERC-ADG 882929 EROS. J.S acknowledges support from the National Research Foundation of Korea (NRF) grant funded by the Korea government (MSIT) (No. 2022R1C1C1005970). S.F. acknowledges support from the Rowland Fellowship at the Rowland

13Institute at Harvard University. S.D.S acknowledges the EPSRC (EP/R023980/1, EP/012932/1 and EP/S030638/1) for funding.

**Competing interests**: The authors declare that they have no competing interests.

**Data and materials availability**: The data that support the plots within this paper and other findings of this study are available at the University of Cambridge Repository (https://doi.org/XXXXX).

**Rights Retention Statement:** This work was funded the UKRI. For the purpose of open access, the author has applied a Creative Commons Attribution (CC BY) license to any Author Accepted Manuscript version arising.13